\documentclass[showpacs,amsmath,showkeys]{revtex4}
\usepackage{graphicx}
\usepackage{graphics}
\usepackage{epsfig}
\usepackage{amsmath}
\begin{document}
\title{ Embedded soliton solutions : A variational study}
\author{Debabrata Pal , Sk. Golam Ali and B. Talukdar }
\email{binoy123@bsnl.in}
\affiliation{Department of Physics, Visva-Bharati University,
Santiniketan 731235, India}
\begin{abstract}
We use a variational method to construct soliton solutions for systems characterized by opposing dispersion and competing  nonlinearities at fundamental and second harmonics. We show that both ordinary and embedded solitons tend to gain energy when the second harmonic field becomes weaker than the first harmonic field.
\end{abstract}
\maketitle
\section{Introduction}
Embedded solitons (ES) represent solitary waves which reside inside the continuous spectrum of a nonlinear - wave system. This type of solitons was first reported by Yang et. al. [1] in optical models characterized by opposing dispersion and competing nonlinearities at fundamental and second harmonics. More specifically , optical media with quadratic $\chi^{(2)}$ and cubic $\chi^{(3)}$ nonlinear susceptibilities can support ES solutions.  The evolution of ES is governed by the coupled nonlinear partial differential equations [2] 
\begin{equation}
iu_z+\frac{1}{2}u_{2t}+u^{*}v+\gamma_1|u|^{2}u+4\gamma_2\mid v\mid^2 u =0 
\end{equation}
and 
\begin{equation}
 iv_z-\frac{1}{2}\delta v_{2t}+qv+\frac{1}{2} u^2+2\gamma_2(\mid v\mid^2 +2\mid u\mid^2 )v=0 , 
\end{equation}
where $u=u(z,t)$ and $v=v(z,t)$ represent the fundamental harmonic (FH) and second harmonic (SH) fields respectively. In writing $(1)$ and $(2)$ we have used  optical notations such that $z$ and $t$ stand for the propagation distance and reduced time. The quantity $-\delta$ is the relative dispersion of SH  and $\gamma_{1,2}$ are the Kerr coefficients.Here q represents the group velocity mismatch originated by the frequency difference of FH and SH fields.
\par
Usually, ES's are studied  using numerical routines to solve $(1)$ and $(2)$. In view of this one often works within the framework of a simplified physical model  where $|v|^2\ll|u|^2$ and neglects the cross-phase modulation (XPM) term (fifth term) in comparison with self-phase modulation (SPM) (fourth term) in $(1)$ .  The SPM term in $(2)$ is also assumed to be negligible in comparison with its XPM counterpart.  Thus we get  a truncated model represented by
\begin{equation}
 iu_z+\frac{1}{2}u_{2t}+u^{*}v+\gamma_1|u|^{2}u =0 
\end{equation}
and 
\begin{equation}
iv_z-\frac{1}{2}\delta v_{2t}+qv+\frac{1}{2} u^2+4\gamma_2 \mid u\mid^2 v=0. 
\end{equation}
For stationary soliton solutions one can use
\begin{equation}
u(z,t)=e^{ikz}U(t),\,\,\,\,\,\,\,\,v(z,t)=e^{2ikz}V(t) 
\end{equation}
with $k$, the FH wave number. The partial differential equations of the full model and those of the truncated model then reduce to ordinary differential equations given by
\begin{equation}
-kU+\frac{1}{2}\ddot{U} +UV+\gamma_1 U^3+4\gamma_2 V^2U=0, 
\end{equation}
\begin{equation}
-2kV-\frac{1}{2}\delta \ddot{V}+qV+\frac{1}{2}U^2+2\gamma_2 (V^2+2U^2)V=0
\end{equation}
and 
\begin{equation}
-kU+\frac{1}{2}\ddot{U}+UV+\gamma_1U^3=0,
\end{equation}
\begin{equation}
-2kV-\frac{1}{2}\delta \ddot{V}+qV+\frac{1}{2}U^2+4\gamma_2 U^2V=0.
\end{equation}
Here the dots denote differentiation with respect to $t$ . 
 Linearization of the equations in $(1)$ and $(2)$ [Full model] as well as in $(3)$ and $(4)$ [Truncated model] shows that both models support ordinary soliton solutions in the regions 
\begin{equation}
0< k < \frac{q}{2} \,\,\,{\rm{if}}\,\,\, \delta > 0, \,\,\,k> {\rm{max}}\left\lbrace 0, \frac{q}{2}\right\rbrace \,\,\,{\rm{if}}\,\,\,\delta < 0 
\end{equation}
and embedded soliton soliton solutions in the regions 
\begin{equation}k> {\rm{max}}\left\lbrace 0, \frac{q}{2}\right\rbrace \,\,\,{\rm{if}} \,\,\,\delta > 0,\,\,\,0< k < \frac{q}{2} \,\,\,{\rm{if}}\,\,\, \delta > 0 . 
\end{equation}
\par
The object of the present work is to derive a straightforward analytical model for comparing the properties of soliton solutions supported by the pair of equations representing the full and truncated models .In doing so we shall consider the cases of ordinary and embedded solitons separately .  To achieve this we shall envisage a variational approach to the problem, where one begins with a Lagrangian  for the system under consideration and constructs the so-called effective Lagrangian by taking recourse to the use of trial  functions for the field variables. Understandably , the trial functions will involve a number of unknown parameters . As we shall see the effective Lagrangian will provide a natural basis to determine these parameters .
\par 
In the above context we note that $(6)$ and $(7)$ , resulting from the full model , follow from an action principle . In contrast to this , $(8)$ and  $(9)$ pertaining to the truncated model are non-Lagrangian . But the latter set of equations are based on physically founded assumptions . This led Kaup and Malomed [2] to adapt the variational approach to the seemingly flawed system represented by $(8)$ and $(9)$ . In their method one starts with the Lagrangian of the full system and drops the term containing $V^4$ to construct an expression for the effective Lagrangian by using the trial functions for $U$ and $V$ . Further ,  the implementation of the Ritz optimization procedure to evaluate  the variational parameters requires one more approximation . We claim that the results in Ref (2) can be rederived and reexamined without taking recourse to the use of this two - tier approximation .In particular , we find that if we work with the effective Lagrangian of the full system , construct equations for the variational parameters and then use the approximation $V\ll U$ ,we automatically arrive at the results of Kaup and Malomed . More significantly , the method followed by us provides a natural basis to examine how the results for $U$ and $V$ for the full model differ from those of the truncated model . One of our main objectives in this work is to compare the results of the full and truncated models and thereby gain some physical weight for the problem .
\par We begin section II with the Lagrangian of the full system and construct the expression for the effective Lagrangian using some trial functions for $U$ and $V$. We then apply the Ritz optimization procedure to obtain equations for the parameters of the trial  functions and examine how the results of Ref. 2 are obtained for $V\ll U$ . In section III we compare the results of $U$  and $V$ for the full model with those for truncated model. We represent the results for both ordinary and embedded solitons.
{\section{ Variational formulation of $(6)$,$(7)$,$(8)$ and $(9)$}}
\subsection{ Lagrangian representation}
\par Our analysis for the properties of ordinary and embedded soliton solutions supported by the full and truncated models will involve essentially a Ritz optimization procedure [3] based on the variational functional for $(6)$ and $(7)$ . It is easily seen that these initial-boundary value problems can be converted to a variational problem with the Lagrangian written as
\begin{eqnarray}
 L = \int{\bigg\lgroup} -kU^2-(2k-q)V^2-\frac{1}{2}\dot{U}^2+\frac{\delta}{2}\dot{V}^2+U^2V+\nonumber\\\frac{\gamma_1}{2}U^4+ 4\gamma_2U^2V^2+\gamma_2V^4{\bigg\rgroup} dt.\,\,\,\,\,\,\,\,
\end{eqnarray}
In the Ritz optimization procedure, the first variation of the variational functional is made to vanish within a  set of suitable chosen trial functions. We thus introduce the ansatz [2] 
\begin{equation}
U=A sech (\sqrt{2 k}t)\,\,\,{\rm{ and }}\,\,\,\,V=B sech^2 (\sqrt{2 k}t)
\end{equation}
for the time - dependent parts of the FH and SH fields. Here the amplitudes $A$ and $B$  are variational parameters. The inverse width  $\sqrt{2 k}$ will, however, not be varied. Inserting $(13)$ in $(12)$ and carrying out the time integral we obtain
\begin{eqnarray}
\left\langle L\right\rangle =\frac{2}{3 \sqrt{2 k}} {\bigg\lgroup}-4 k A^2-2 (2 k-q) B^2+\frac{8}{5}\delta k B^2+2 A^2 B+\nonumber\\\gamma_1 A^4+\frac{32}{5}\gamma_2 A^2 B^2+\frac{48}{35}\gamma_2 B^4{\bigg\rgroup},\,\,\,\,\,\,
\end{eqnarray}
the effective Lagrangian for $U$ and $V$ in $(13)$ .  The Lagrangian in $(14)$ represents a specific function of the parameters only. Optimization with respect to parameters will yield a system of equations which when solved will determine $U$ and $V$ within the chosen set of trial functions and a concomitant approximation for the true solutions. This is the route we follow to determine the values of the parameters $A$ and $B$. 
\subsection{ Variational parameters and truncated model}
\begin{figure}
\includegraphics[width=0.4\columnwidth, angle=-90] {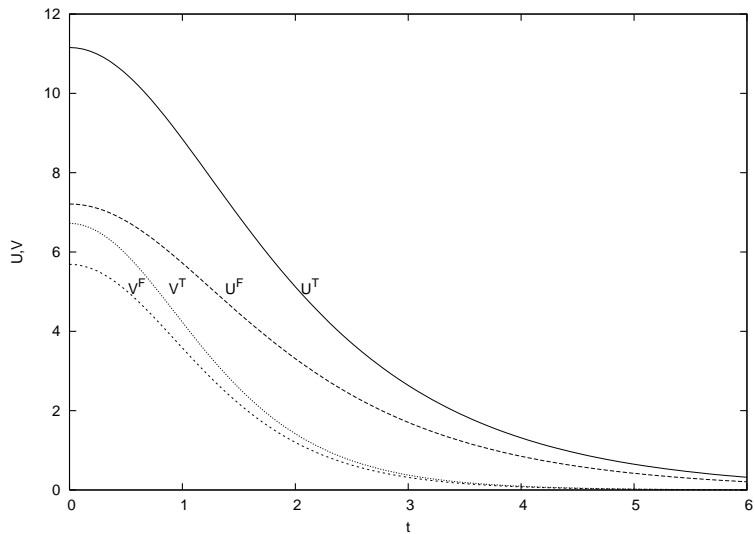}
\caption{ $U$ and $V$ as a function of $t$ for non - embedded solitons.}
\end{figure}
\begin{figure}
\includegraphics[width=0.40\columnwidth, angle=-90] {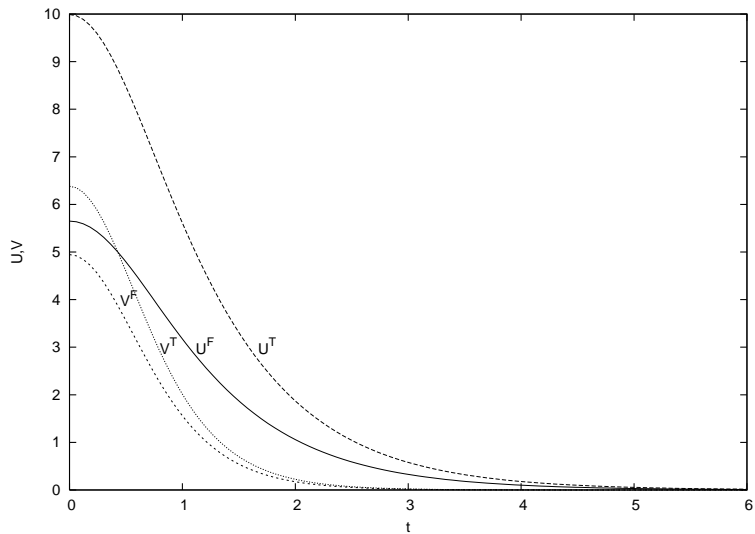}
\caption{$U$ and $V$ as a function of $t$ for embedded solitons.}
\end{figure}
From the vanishing conditions of $\frac{\delta \left\langle L\right\rangle }{\delta A}$ and $\frac{\delta \left\langle L\right\rangle }{\delta B}$  we obtain 
\begin{equation}
- 2k+B+\gamma_1 A^2+\frac{16}{5} \gamma_2 B^2=0 
\end{equation}
and
\begin{equation}
A^2+\frac{32}{5}\gamma_2 A^2 B-2(2 k-q)B+\frac{8}{5}\delta k B+\frac{96}{35}\gamma_2 B^3=0 .
\end{equation}
Understandably, these equations determine the parameters of the full model. To go over to the truncated model we can choose $B\ll A$ and neglect $B^2$ and $B^3$ in $(15)$ and $(16)$ to get 
\begin{equation}
- 2k+B+\gamma_1 A^2=0 
\end{equation}
and
\begin{equation}
A^2+\frac{32}{5}\gamma_2 A^2 B-2(2 k-q)B+\frac{8}{5}\delta kB=0.
\end{equation}
These equations were obtained by Kaup and Malomed [2] first by neglecting the last term in $(12)$ and then  again neglecting the contribution of the term $4\gamma_2 U^2 V^2$ while taking variation with respect to $A$. But we have shown that this type of two - tier approximation is not essential to make a transition from the full to the truncated model.
\section{ Soliton solutions}
We have seen that when the wave number $k$ falls into the region $(10)$ both full and truncated models have ordinary soliton solutions . To see how , in this case ,the results of $U$ and $V$ for the full model differ from those of the truncated model we have chosen to work with $k=0.25,\gamma_1=-0.05,\gamma_2=-0.025,\delta=1,q=1$ . As for the full model , we use these values in $(15)$ and $(16)$ to get three values for $B$ , namely $B_1=0.4824 , B_2=5.6895 , B_3=19.4379$ . We find that $A$ values corresponding to $B_1$ and $B_3$ are imaginary while $A$ becomes a real number equal to $7.2109 (A_2^{F})$ when calculated by using the value of $B_2 (B_2^{F})$ .The corresponding results for the truncated model are $B_2^{T}=6.7227 $ and $A_2^{T}=11.1559$ . The superscripts $F$ and $T$ refer to the full and truncated models . We shall also use similar superscripts on $U$ and $V$ .The ordinary or non-embedded soliton solutions are shown in Fig. $1$ . From this figure it is clear that the curves for $U^{T}$ and $V^{T}$ are more peaked compare to the curves for $U^{F}$ and $V^{F}$ . It will , therefore be interesting to examine how the behaviour of $U^{F} , V^{F} , U^{T}$ and $V^{T}$ is affected in the case of embedded solitons . 
\par
In consistent with $(11)$ we take $k=0.6963,\gamma_1=-0.05,\gamma_2=-0.025,\delta=1,q=1$  for the embedded soliton . In this case we find $B_2^{F}=4.9450 , A_2^{F}=5.6500 , B_2^{T}=6.3822$ and $A_2^{T}=9.9896$ . In Fig. $2$ we display the curves for embedded solitons . In this case also the curves for $U^{T}$ and $V^{T}$ are more peaked than the curves for $U^{F}$ and $V^{F}$ . But looking closely into the curves in Figs. $1$ and $2$ we see that in the case of embedded solitons the curves for $U^{T}$ and $V^{T}$ fall off more rapidly than their non-embedded counterparts . 
\par 
In the model considered in this work the energy of the soliton is given by $E =\int_{-\infty}^{+\infty}\left( |u|^2+2 |v|^2\right) dt$. From the results in Figs. 1 and 2 it is clear that $E^T >E^F$  for both ordinary and embedded  solitons. Understandably, $E^T$ and $E^F$ stand for the soliton energies obtained by using the truncated and full models. It will, therefore, be an interesting curiosity to verify how the approximation  $V\ll U$ affects a typical experiment.
\section*{ACKNOWLEDGEMENTS}
One of the authors (BT) would like to  acknowledge the financial support of the University Grants Commission, Government of India (F. No. 32 - 39 / 2006 (SR)). The authors are thankful to Prof. S. N. Roy, Department of Physics, Visva - Bharati, Santiniketan 731235, India


\begin{thebibliography}{99}  
\bibitem{1}  J  Yang , B  A  Malomed and D  J  Kaup,Embedded solitons in Second-Harmonic Generating systems , Phys. Rev. Lett. {\bf 83 }, 1958 (1999)
\bibitem{2} D  J  Kaup, B  A  Malomed , Embedded solitons in Lagrangian and semi-Lagrangian systems , Physica D {\bf 184} 153 (2003)
\bibitem{3} D  Anderson , Variational approach to nonlinear pulse propagation in optical fibers ,  Phys. Rev. A {\bf 27} 3135 (1983)
\end{thebibliography}
\end{document}